\documentclass[12pt]{article}
\usepackage{setspace}
\usepackage[colorlinks=true]{hyperref}
\usepackage{graphicx}
\usepackage{subfigure}
\usepackage{path}
\usepackage{fancyhdr}

\title{A New Efficient Protocol for k-out-of-n Oblivious Transfer}
\author{Ashwin Jain, C Hari
\thanks{Ashwin Jain and C Hari are with the Dhirubhai Ambani Institute of Information and Communication Technology (DA-IICT), Gandhinagar 382007, Gujarat, India.\newline E-mails: \{ashwinrkjain, hari.daiict\} @gmail.com}}

\renewcommand\headheight{24pt}

\begin{document}
\maketitle

\begin{abstract}
This paper presents a new efficient protocol for k-out-of-n oblivious transfer which is a generalization of Parakh's 1-out-of-2 oblivious transfer protocol based on Diffie-Hellman key exchange. In the proposed protocol, the parties involved generate Diffie-Hellman keys obliviously and then use them for oblivious transfer of secrets.
\end{abstract}


\section{Introduction}
Oblivious Transfer ~\cite{cryptoeprint:2005:187, 1451166, 1392816} of secrets between two parties is a very useful primitive for the construction of larger cryptographic schemes. It is a method by which a commodity from a set is transferred from a sender to a receiver based on the receiver's choice. However, the sender should be oblivious to the choice that the receiver made, i.e. he should be unaware of which commodity the receiver is in possession of at the end of the transaction. Oblivious Transfer has applications in the areas of secure multiparty computation, private information retrieval (PIR), fair electronic contract signing, oblivious secure computation, etc. \cite{Goldreich97howto, Aiello01pricedoblivious, 62213, 3818}.

In this paper, we present a $k$-out-of-$n$ generalization of the $1$-out-of-$2$ oblivious transfer protocol proposed by Parakh ~\cite{1451166}. He presented a protocol that established an oblivious key exchange between two parties using the Diffie-Hellman protocol at its core. Once the keys were exchanged the parties would use a symmetric key cryptosystem for the transfer of secret messages, thus making the transfer more efficient compared to using a public key cryptosystem. The scheme may further be used to establish oblivious transfer channel for the transfer of large secrets.

A $k$-out-of-$n$ Oblivious Transfer is when the receiver can choose to receive $k$ secrets from a set of $n$ secrets that the sender is in possession of. For example, Bob may have a set of $n$ files protected by individual passwords that are immune to trial-and-error (due to their length or complexity or both). Alice is in possession of the passwords for these files. Now, Bob wants to open $k$ of these files for which he would need their respective passwords from Alice. Also, he doesn't want Alice to know which of the $n$ files he wishes to read. Oblivious Transfer can come to the rescue in such a situation. It will enable Bob to learn the passwords of the $k$ files that he wants to read and at the same time, prevent Alice from knowing which passwords Bob has actually acquired. One must also bear in mind that given the $k$ passwords, it should not be possible for Bob to compute any of the remaining $(n-k)$ passwords.\\

Thus, the goals of Oblivious Transfer can be summarized as follows:
\begin{itemize}
 \item {\bf Receiver's Privacy:} Alice should not be able to determine which $k$ secrets Bob has acquired.
 \item {\bf Sender's Privacy:}  Bob should not be able to learn any of the remaining $(n - k)$ secrets using the $k$ secrets that he has received.
\end{itemize}

\section{Previous Work}
\textit{Rabin's} Oblivious Transfer protocol allowed the receiver to receive a bit with a probability $\frac{1}{2}$. The sender on the other hand, could not determine whether the receiver has received the bit or not. This idea was later used to establish 1-out-of-2 OT protocols that can be extended easily to 1-out-of-$n$ protocols ~\cite{1382945} and these in turn can be converted into $k$-out-of-$n$ protocols by merely running the protocol $k$ times ~\cite{Tzeng02efficient1-out-n}. However, as expected, the computational cost of these extended protocols would be high. It is possible to reduce the complexity by developing 1-out-of-$n$ and $k$-out-of-$n$ protocols directly from primitives (without the successive runs of lower order protocols) ~\cite{1382945, DBLP:journals/eatcs/SalomaaS90, 716710}. Both the possibilities of successive protocol runs and direct implementation have been explored in Oblivious Transfer protocols ~\cite{Cachin98onthe}.

In ~\cite{Chu:jucs_14_3:efficient_k_out_of}, \textit {Chu and Tzeng} devised a scheme for implementation of 1-out-of-$n$ and $k$-out-of-$n$ protocols based on the Discrete Log problem. They compared the cost of their protocol to that of \textit{Mu, Zhang, and Varadharajan} ~\cite{678314} and \textit {Naor and Pinkas} ~\cite{Naor99oblivioustransfer}. Although their 1-out-of-$n$ protocol was of $O(n)$, their $k$-out-of-$n$ protocol used $k$ successive runs of their 1-out-of-$n$ protocol. This increases the cost of their $k$-out-of-$n$ scheme to $O(kn)$. \textit{Wu, Zhang, and Wang} ~\cite{WuZW03} improved this efficiency in their paper and developed a protocol that was of $O(k+t)$ using a two lock cryptosystem. This protocol does not involve the use of \textit{Diffie-Hellman} based keys. An efficient oblivious transfer protocol using Elliptic Curve Cryptography was presented in \cite{1392816}.

\section{Parakh's Oblivious Transfer Protocol}

Oblivious transfer using Diffie-Hellman keys was presented in \cite{1451166} . Here, Alice encrypts the two secrets she is willing to disclose, under two different encryption keys and associates these keys with two distinct choices. She then establishes a $1$-out-of-$2$ oblivious key exchange such that Bob is able to only compute one of the keys based on his choice. Consequently, upon receiving the encrypted secrets, Bob is only able to decrypt one of them.

We provide a brief description of the protocol here in order to make the idea of oblivious key exchange clear. However, our description differs slightly from that presented in \cite{1451166} because we note that the pre-requisite of choosing two numbers $x_1$ and $x_2$ such that $c=x_1^2 =x_2^2 \pmod {p}$ is not necessary for successful execution of the protocol.

Assuming a safe prime $p$, a generator $g$, and $x_1$ and $x_2$ be two randomly and uniformly chosen numbers from the field $Z_p$, denote the two secrets that Alice possesses by $S_1$ and $S_2$. She then associates $x_1$ with $S_1$ and $x_2$ with $S_2$ (without disclosing the secrets). She announces these associations to Bob; denote Bob's choice by $x_B$. Bob's task is to establish either key $K_1$ or $K_2$ with Alice, according to which secret he is interested in obtaining.

The protocol proceeds as follows:
\begin{enumerate}
 \item Alice secretly chooses $N_{A_1}$ and sends to Bob: $g^{x_1 + N_{A_1}} \pmod {p}$;
 \item Bob chooses $x_B$ = $x_1$ (if he wants secret $S_1$) or $x_B = x_2$ (if he wants secret $S_2$) and secret numbers $N_B$ and $N_{B_1}$;
 \item Bob sends to alice: $\Big(\frac{g^{x_1 + N_{A_1}}}{g^{x_B}}\Big)^{N_BN_{B_1}} \pmod {p}$ and $g^{N_B} \pmod {p}$;
 \item Alice chooses a number $N_{A_2}$ and sends to Bob: $\Big[\Big(\frac{g^{x_1 + N_{A_1}}}{g^{x_B}}\Big)^{N_BN_{B_1}}\Big]^{N_{A_2}} \pmod {p}$;
 \item Bob computes: $K_B \equiv \Big[\Big(\frac{g^{x_1 + N_{A_1}}}{g^{x_B}}\Big)^{N_BN_{B_1}N_{A_2}}\Big]^{\frac{1}{N_{B_1}}} \pmod {p}$ $\equiv \Big(\frac{g^{x_1 + N_{A_1}}}{g^{x_B}}\Big)^{N_BN_{A_2}} \pmod {p}$;
 \item Alice computes: $K_1 \equiv g^{N_BN_{A_1}N_{A_2}} \pmod {p}$ and $K_2 \equiv (g^{N_B(x_1-x_2+N_{A_1})})^{N_{A_2}} \pmod {p}$; and
 \item Alice encrypts secret $S_1$ using $K_1$ and secret $S_2$ using $K_2$ and sends them to Bob.
\end{enumerate}

From the above sequence we see that if Bob chooses $x_B = x_1$, then $K_B = K_1$, and if Bob chooses $x_B = x_2$, then $K_B = K_2$. Hence, Bob will only be able to retrieve one of the two secrets depending upon his choice, while Alice will not be able to determine which secret Bob has retrieved. Hence, Bob has obliviously established a secret key, or his choice, with Alice.

\section{Assumptions in this Paper}
Throughout the paper we assume that Alice is the party having possession of $n$ secrets or in other words, is the sender. Bob is the party that wants to learn one or more secrets obliviously. Alice and Bob are both assumed to be honest but curious parties, i.e. in spite of their honesty, they will try to obtain more information than they are entitled to.

The protocol has no way assuring the legitimacy of the secrets handed over by Alice to Bob during the transaction. However, for the purpose of this protocol we do assume that any message exchange between two parties over a channel is duly signed by the sender. In case of a fraud (in the contents of the messages) the victim can later use these digital signatures as evidence against the adversary during adjudication.

\section{1-out-of-n Oblivious Transfer}

For the security of the protocol, we have exploited the fact that finding the exponent $e$ in the equation $x^e \pmod{p} = y$ where $x$ and $y$ are given) is equivalent to solving a discrete log problem (DLP). Let $g \in Z_p$ be the generator of the \textit{Diffe-Hellman} group $Z_p$ where $p$ is considered to be a safe prime.

Let there be a set of numbers $x_1, x_2, ... , x_n$ known both to Alice and Bob. Say Alice has $n$ secrets
$S_1, S_2, ... , S_n$ and Bob wants to acquire the $i^{th}$ secret $S_i$, then Bob will choose $x_i$ for the generation of key as per the protocol.

Let $K_{A_i}$ be the key used by Alice to encrypt the secret $S_i$ for all $i$, and $K_B$ be the key generated by the Bob for decryption of the secret. $N_{A_1}$ and $N_{A_2}$ are ephemeral nonces generated by Alice and $N_{B_1}$, $N_{B_2}$ and $N_{B_3}$ are the ephemeral nonces generated by Bob in the protocol run.

\subsection{Mutual Agreement}
Alice and Bob both agree upon a safe prime $p$, a generator element $g$ of group $Z_p$ and the set $\{x_1, x_2, ..., x_{m-1}, x_m\}$. Each member $x_i$ of the set corresponds to the $i^{th}$ secret. All the nonces generated by the parties are ephemeral.

\subsection{The Protocol}
\begin{enumerate}
 \item Alice generates random nonce $N_{A_1}$ and sends the message $M_A  = g^{N_{A_1}+\Sigma_{i=1}^{n}x_i} \pmod{p}$ to Bob.
 \item Bob selects $x_j$ as per the secret he wants to acquire, and generates three nonces $N_{B_1}$, $N_{B_2}$ and $N_{B_3}$ such that $N_{B_3} = k$ x $N_{B_2}$ where $k$ is a factor of $N_{B_1}$.
 \item Bob sends the message \newline \newline$M_1 = (\frac{M_A}{g^{x_j} \pmod{p}})^{\frac{N_{B_1}N_{B_2}}{N_{B_3}}} \pmod {p}$ to Alice.
 \item Bob also sends $M_B  = g^{N_{B_1}} \pmod {p}$.
 \item Alice generates nonce $N_{A_2}$ and the set of keys $\{K_{A_1}, K_{A_2}, ..., K_{A_{n-1}}, K_{A_n}\}$ as \newline $K_{A_k} = ({(M_B)}^{N_{A_1}+\Sigma_{i=1}^{n}x_i - x_k})^{N_{A_2}} \pmod {p} \forall k \in [1, n]$.\newline
 \item Alice sends the message $[M_1]^{N_{A_2}} \pmod {p}$ to Bob.\newline
 \item Bob calculates $K_B$ as $[[M_1]^{N_{A_2}}]^{\frac{N_{B_3}}{N_{B_2}}} \pmod {p}$.
 \item Alice sends all the secrets encrypted under the respective key ($S_i$ is encrypted under the key generated $K_{A_i}$), i.e. $\{S_1\}_{K_{A_1}}$, $\{S_2\}_{K_{A_2}}$, $\{S_3\}_{K_{A_3}}$,... $\{S_n\}_{K_{A_n}}$.
 \item Bob can then decrypt the locked secret that he wished to learn using the key $K_B$ he has generated.
\end{enumerate}

\begin{figure*}[ht]
\centering
\includegraphics[scale=0.5]{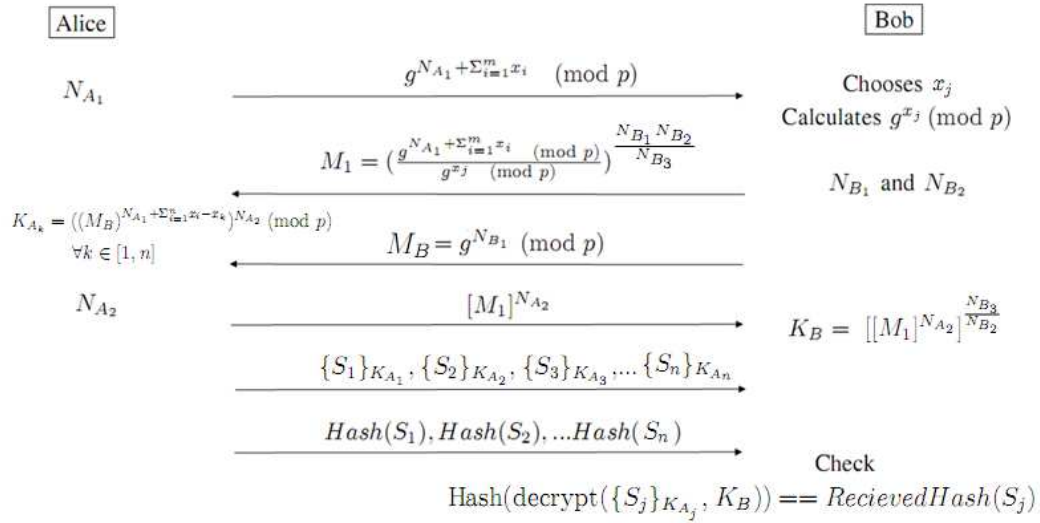}
  \caption{1-out-of-n Oblivious Transfer Protocol Run}
\label{protocol}
\end{figure*}

\subsection{Security Proof and Cost Analysis}
It is easy to see that if Alice wishes to know Bob's choices she would have to know $x_i$ that is conveyed in the form $g^{x_i} \pmod {p}$. In order to do this, she would have to solve the Discrete Log Problem. However, solving the Discrete Log Problem is considered computationally intractable. Thus, receiver's privacy is assured.

If Bob wishes to acquire more than the $k$ secrets he is entitled to, he will have to obtain the nonce $N_{A_2}$ which is again equivalent to solving the Discrete Log Problem, thus ascertaining sender's privacy.	

The computational costs due to exponentiation at Alice's and Bob's ends are $n+1$ and 2 i.e. $O(n)$ and $O(1)$ respectively. The transfer cost is quite plainly $n+4$ i.e. $O(n)$. This is equal in order to the protocol proposed in ~\cite{Chu:jucs_14_3:efficient_k_out_of} which is also based directly on cryptographic primitives.

\subsection{Same Message Attack}
However, the protocol is vulnerable against the same message attack. i.e. if all the secrets that Alice sends are the same, then (trivially) no matter which secret Bob chooses, Alice will always know the secret he has chosen. This attack can be avoided with a simple addition of the following steps to the protocol.

\begin{enumerate}
 \item Alice also sends the hash value of each secret to Bob that is $Hash(S_1)$, $Hash(S_2)$, ... $Hash(S_n)$.
 \item Bob verifies if all the hash values received are distinct. If Alice has sent distinct secrets and hashed them honestly, then the hashes will prove to be different.
 \item Bob then decrypts $\{S_{A_i}\}_{K_i}$ using $K_{B_i}$ calculated by him.
 \item Check if \newline Hash(decrypt($\{S_j\}_{K_{A_j}}$, $K_B$)) == $RecievedHash(S_j)$. In case the match fails, it means that Alice has either sent him fake hashes in order to make them different, or she has hashed them dishonestly.
\end{enumerate}

Alice will have an extremely low probability of getting away with a Same Message Attack. It will happen only in the case that Alice hashes only one secret honestly, fakes the other hashes and Bob picks the secret that is hashed correctly. We assume that the probability of this happening will be very low.

\section{k-out-of-n Oblivious Transfer}
$k$-out-of-$n$ Oblivious Transfer scheme is when Alice is in possession of $n$ secrets and Bob wishes to learn $k$ of them. This can, of course, be achieved by running our 1-out-of-$n$ protocol $k$ times, once for each secret. But, it would save computation and transfer cost if we establish a different protocol for the same that is inspired from our $1$-out-of-$n$ protocol. The proposed $k$-out-of-$n$ protocol is again reliant on the Discrete Log Problem for its security and uses \textit{Diffie-Hellman} ~\cite{diffie76new} based keys for locking and unlocking secrets.

\subsection{Mutual Agreement}
Alice and Bob both agree upon a safe prime $p$, a generator element $g$ of group $Z_p$ and the set $x_1, x_2, ... x_n$. Each member $x_i$ of the set corresponds to the $i$th secret. They also agree upon the number of secrets to be transferred $k$.

\subsection{The Protocol}
\begin{enumerate}
 \item Alice generates random nonce $N_{A_1}$ and sends the message $M_A =g^{N_{A_1}+\Sigma_{i=1}^{n}x_i} \pmod{p}$ to Bob.
 \item Bob selects $\{x_1, x_2, ... x_k\}$ as per the secrets he wants to acquire, and generates three nonces $N_{B_1}$, $N_{B_2}$ and $N_{B_3}$ such that $N_{B_3} = k$ x $N_{B_2}$ where $k$ is a factor of $N_{B_1}$.
 \item Bob sends the messages \newline \newline$M_j = (\frac{M_A}{g^{x_j} \pmod{p}})^{\frac{N_{B_1}N_{B_2}}{N_{B_3}}} \pmod{p}\forall j \in [1,k]$ to Alice.
 \item Bob also sends $M_B  = g^{N_{B_1}} \pmod {p}$.
 \item Alice generates nonce $N_{A_2}$ and the set of keys $\{K_{A_1}, K_{A_2}, ..., K_{A_{n-1}}, K_{A_n}\}$ as \newline $K_{A_j} = ({(M_B)}^{N_{A_1}+\Sigma_{i=1}^{n}x_i - x_j})^{N_{A_2}} \pmod{p}\forall j \in [1, n]$.\newline
 \item Alice sends the messages $[M_j]^{N_{A_2}} \pmod{p}\forall j \in [1,k]$ to Bob.\newline
 \item Bob calculates $K_{B_j}$ as $[[M_j]^{N_{A_2}}]^{\frac{N_{B_3}}{N_{B_2}}} \pmod{p}\forall j \in [1,k]$.
 \item Alice sends all the secrets encrypted under the respective key ($S_i$ is encrypted under the key generated $K_{A_i}$), i.e. $\{S_1\}_{K_{A_1}}$, $\{S_2\}_{K_{A_2}}$, $\{S_3\}_{K_{A_3}}$,... $\{S_n\}_{K_{A_n}}$.
 \item Bob can then decrypt the locked secrets that he wished to learn using the keys $K_{B_j}, \forall j \in [1,k]$ he has generated.
\end{enumerate}

Let us understand the working of the above protocol with an example.\\

\noindent \textbf{Example:} Alice is in possession of say 5 secrets, $S_1, S_2, S_3, S_4, S_5$ (i.e. n=5). They agree upon the safe prime $p=23$, the generator  $g=5$ of the group $Z_{23}$ and the set $\{1,2,3,4,5\}$ such that ${1}$ corresponds to $S_1$, ${2}$ corresponds to $S_2$ and so on. They also decide the number of secrets to be transferred $k=2$.

\begin{enumerate}
\item Alice generates nonce $N_{A_1} = 4$ and sends \newline
\begin{tabular}{ r l }
 $M_A = 5^{4 + (1+2+3+4+5)} \pmod {23} $&$ \equiv 5^{19} \pmod{23}$ \\
 & $\equiv 7$
\end{tabular}
\item Suppose Bob wants secrets $S_3$ and $S_5$. He therefore chooses $x_1  = 3$ and $x_2 = 5$. He generates the nonces $N_{B_1} = 10$, $N_{B_2} = 6$ and $N_{B_3} = 12$. [Here, $N_{B_3} = k$ x $N_{B_2}$ where $k = 2$ which is a factor of $N_{B_1}$].
\item Bob calculates and sends the messages \newline
\begin{tabular}{ r l }
 $M_1 = {(\frac{7}{10})}^{5} \pmod{23}$ & $\equiv ({7 \times 10^{-1}})^5\pmod {23}$ \\
 & $ \equiv 3^5 \pmod {23}$ \\
 & $\equiv 13$
\end{tabular}\newline
\begin{tabular}{ r l }
 $M_2 = {(\frac{7}{20})}^5 \pmod{23}$&$ \equiv ({7 \times 20^{-1}})^5\pmod{23}$ \\
 & $\equiv 13^5 \pmod {23}$ \\
 & $\equiv 4$
\end{tabular}
\item Bob also sends $M_B = 5^{10} \pmod {23} \equiv 9$.
\item Alice generates nonce $N_{A_2} = 8$ and the calculates the following keys:\newline
$K_{A_1} = (9 ^{19-1})^8 \pmod {23} \equiv 9$\newline
$K_{A_2} = (9 ^{19-2})^8 \pmod {23} \equiv 6$\newline
$K_{A_3} = (9 ^{19-3})^8 \pmod {23} \equiv 4$\newline
$K_{A_4} = (9 ^{19-4})^8 \pmod {23} \equiv 18$\newline
$K_{A_5} = (9 ^{19-5})^8 \pmod {23} \equiv 12$\newline
Alice encrypts $S_1$ with the key $K_{A_1}$, $S_2$ with the key $K_{A_2}$ and so on.
\item Alice calculates and sends ${M_1}^{N_{A_2}} \pmod {p} = 13^8 \pmod {23} \equiv 2$ and ${M_2}^{N_{A_2}} \pmod {p} = 4^8 \pmod {23} \equiv 9$ to Bob.
\item Bob calculates $K_{B_1} = 2^{\frac{12}{6}} \pmod {23} \equiv 4$, and \newline$K_{B_2} = 9^{\frac{12}{6}} \pmod {23}\equiv 12$.
\item Alice sends all the encrypted secrets to Bob i.e. ${\{S_1\}}_{K_{A_1}}$, ${\{S_2\}}_{K_{A_2}}$, ${\{S_3\}}_{K_{A_3}}$, ${\{S_4\}}_{K_{A_4}}$ and  ${\{S_5\}}_{K_{A_5}}$.
\item We can see that the keys generated for $S_3$ and $S_5$ by both Alice and Bob are $4$ and $12$ respectively.
\end{enumerate}

Thus, the generated keys by Alice and Bob (i.e. $K_{A_j}$ and $K_{B_j}$) for all the chosen secrets ([1…k]) are the same. The keys have thus been exchanged by parties obliviously and can use a symmetric key cryptosystem for the transfer of secrets.

\subsection{Cost Analysis}
The computational cost at Alice's and Bob's end can be seen to be $n+k$ and $2k$ respectively [$O(n+k)$ and $O(2k)$]. This is equal to the computational cost at either end in the scheme proposed in ~\cite{WuZW03}. The transfer cost would be equal to $n+2k+2$ [$O(n+k)$]. This again is equal in order to the scheme proposed in the paper in ~\cite{WuZW03}.

\section{Conclusion}
The protocol in this paper equals the order of the 1-out-of-n protocol in ~\cite{Chu:jucs_14_3:efficient_k_out_of} both in computation and transfer. For $k$-out-of-$n$ Oblivious Transfer, it compromises on the adaptive nature of their protocol and requires that both parties decide on the number $k$ of secrets to be transferred before the execution of the actual protocol. However, it improves the cost of their $k$-out-of-$n$ protocol and equals the order of the scheme proposed in ~\cite{WuZW03}. The hash function used to avoid the same message attack takes negligible computational cost due to the availability of very fast hashing algorithms. The transfer of these also induces a minor overhead that does not affect the order of the transfer cost.

The protocol uses \textit{Diffie-Hellman} ~\cite{diffie76new} based keys to encrypt and decrypt the secrets. Our scheme basically allows both the parties to obliviously generate \textit{Diffie-Hellman} keys. Such a primitive can be used in other applications that use \textit{Diffie-Hellman} based keys to ensure privacy.

Although the order of the $k$-out-of-$n$ protocol presented in this paper and that proposed in ~\cite{WuZW03} are the same, it is important to note that the all the three rounds in the scheme proposed by \textit{Wu et.al.} ~\cite{WuZW03} involve the transmission of the secret itself in an encrypted form. For smaller secrets, both the protocols may exhibit similar performance. However, as the size of the secrets increases, (in case of files)  ~\cite{WuZW03}'s protocol would have the rather unnecessary overhead of transmitting the entire file in its encrypted form (which of course cannot be significantly smaller than the file itself). Our protocol on the other hand, transmits the encrypted secret only once and thus will save significant bandwidth in a scenario involving large secrets. We believe that such a scenario may occur frequently in applications such as internet shopping for digital commodities, exchange of digital secrets, file transfers, etc.  Our protocol would be able to perform significantly better under such circumstances.

\bibliographystyle{abbrv}
\bibliography{final}

\end{document}